\documentclass[sigconf,authorversion]{acmart}

\copyrightyear{2023}
\acmYear{2023}
\setcopyright{cc}
\setcctype{by}
\acmConference[ARES 2023]{The 18th International Conference on Availability, Reliability and Security}{August 29 -- September 1, 2023}{Benevento, Italy}
\acmBooktitle{The 18th International Conference on Availability, Reliability and Security (ARES 2023), August 29-September 1, 2023, Benevento, Italy}
\acmPrice{15.00}
\acmDOI{10.1145/3600160.3600187}
\acmISBN{979-8-4007-0772-8/23/08}


\usepackage{graphicx}
\graphicspath{ {./images/} }

\usepackage{hyperref} 

\usepackage[left = `,right = ',leftsub = ``,rightsub = '']{dirtytalk}

\newcommand{\aitm}{AitM}
\newcommand{\depend}{\sim}
\newcommand{\sigwhat}{Signal\,/ \!WhatsApp}

\begin{document}
\date{}

\title[The Effect of Length on Key Fingerprint Verification Security and Usability]{The Effect of Length on \protect\\ Key Fingerprint Verification Security and Usability}

\author{Dan Turner}
\orcid{0009-0004-8338-558X}
\affiliation[obeypunctuation=true]{%
    \country{}
    }
\email{dan@turnerhallow.co.uk}
 
\author{Siamak F. Shahandashti}
\orcid{0000-0002-5284-6847}
\affiliation[obeypunctuation=true]{%
    \institution{University of York},
    \country{UK}
    }
\email{siamak.shahandashti@york.ac.uk}

\author{Helen Petrie}
\orcid{0000-0002-0100-9846}
\affiliation[obeypunctuation=true]{%
    \institution{University of York},
    \country{UK}
    }
\email{helen.petrie@york.ac.uk}

\begin{abstract}
In applications such as end-to-end encrypted instant messaging, secure email, and device pairing, users need to compare key fingerprints to detect impersonation and adversary-in-the-middle attacks.
Key fingerprints are usually computed as truncated hashes of each party's view of the channel keys, encoded as an alphanumeric or numeric string, and compared out-of-band, e.g.\ manually, to detect any inconsistencies.
Previous work has extensively studied the usability of various verification strategies and encoding formats, however, the exact effect of key fingerprint length on the security and usability of key fingerprint verification has not been rigorously investigated. 
We present a 162-participant study on the effect of numeric key fingerprint length on comparison time and error rate.
While the results confirm some widely-held intuitions such as general comparison times and errors increasing significantly with length, a closer look reveals interesting nuances.
The significant rise in comparison time only occurs when highly similar fingerprints are compared, and comparison time remains relatively constant otherwise.
On errors, our results clearly distinguish between security non-critical errors that remain low irrespective of length and security critical errors that significantly rise, especially at higher fingerprint lengths.
A noteworthy implication of this latter result is that \sigwhat\ key fingerprints provide a considerably lower level of security than usually assumed.
\end{abstract}


\maketitle

\section{Introduction}
\label{sect:Intro}
Authentic keys are required for secure communication. Devices negotiate these keys using a key exchange protocol, or use a public key that  purportedly belongs to the other party.
These keys may be authenticated using authenticated key exchange protocols such as password-based authenticated key exchange, or by verifying public key certificates.
Such authentication is only possible when there is an existing shared security context between parties such as shared passwords or public key infrastructure (PKI). 
In the absence of authentication, adversaries may carry out adversary-in-the-middle (\aitm, traditionally known as  man-in-the-middle) or impersonation attacks to compromise security.

Digital devices have become ubiquitous, and hence there is a growing need for establishing ad hoc secure communication channels between devices, i.e.\ securely \emph{pairing} devices, without a shared security context.
Although impersonation and \aitm\ attacks cannot be prevented, system designers can build in measures to \emph{restrict} or \emph{detect} such attacks.
As an example of the restriction approach, distance bounding protocols in contactless payment systems limit the distance between the payment card or device and the point of sale terminal to minimise the possibility of \aitm\ attacks~\cite{brands1993distance}, such as the so-called Mafia Fraud.

One of the most common methods to detect impersonation or \aitm\ attacks is through an \emph{out-of-band channel}.
System designers assume that users have access to a separate secure communication channel with low bandwidth.
The key observation is that the keys held by the communicating parties will differ when there is an impersonation or \aitm\ attack, and will be identical in the absence of such attacks.
The out-of-band channel is used to detect differences between the keys the two sides hold after the key exchange.
Since the out-of-band channel is low bandwidth, devices usually apply a hash function to the keys and truncate the result to derive a short digest, which we call a \emph{key fingerprint}.
Comparing the short fingerprints through an out-of-band channel would provide the confidence in keys being identical bar any hash collisions.

Various formats for key fingerprints have been considered.
OpenPGP, designed for email encryption, encodes public-key fingerprints as hexadecimal strings. The user then manually compares these against a trusted copy of the key fingerprint, e.g., on a business card.
The ZRTP protocol for secure VoIP uses a Short Authentication String (SAS), which is a fingerprint of the key negotiated using Diffie--Hellman key exchange.
The Silent Phone app shows the SAS as two words for users to verbally check.
Loud and Clear, a device pairing method, creates a short sentence from the key fingerprint and speaks it aloud using a text-to-speech engine.
The user checks it against a sentence shown on the other device~\cite{goodrich2006loud}. 

\begin{figure}
  \centering
  \fcolorbox{gray}{white}{\includegraphics[scale=0.15]{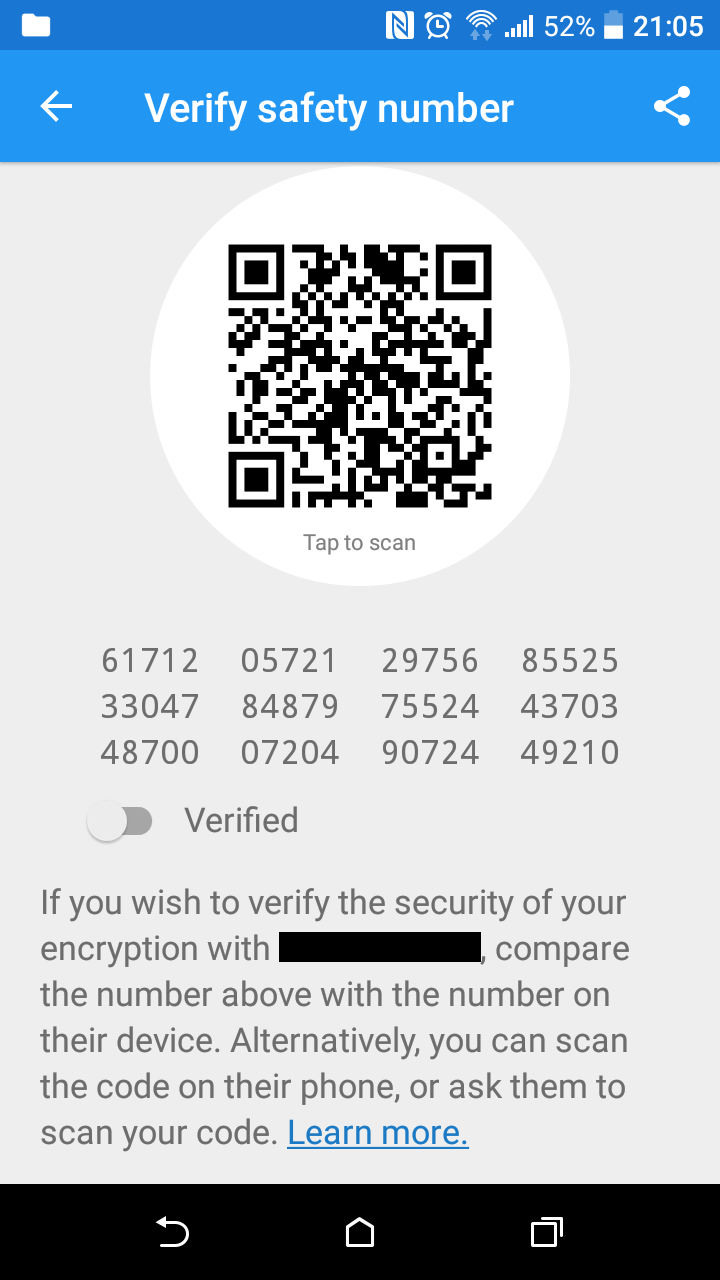}}
  \caption{Safety number display in Signal for Android} 
  \label{img:signal}
\end{figure}

Alphanumeric fingerprints are one of the most widely used as they are generally considered comparatively more usable.
The most widely deployed text-based format is likely to be numeric, thanks to WhatsApp's 2016 rollout of the Signal protocol for end-to-end encryption.
Signal and WhatsApp use a string of 60 digits arranged in 12 chunks of 5 as shown in Figure~\ref{img:signal}.
This is called a \say{safety number} in Signal and a \say{security code} in WhatsApp.

The key fingerprint length is usually set based on the security level required for the application.
Signal and WhatsApp use 60-digit fingerprints since they need to provide long-term security against adversaries without any location restriction.
However, a 4-digit fingerprint may be sufficient to safely pair two smart home devices if keys are freshly generated for one-time use and the communication protocol is short range.

There have been multiple studies on the usability of various key fingerprint formats and their susceptibility to error in the literature.
However, apparently no study has investigated the effect of key fingerprint length. 
Intuitively, one expects that users can compare shorter key fingerprints more quickly and with fewer errors, but the veracity of this intuition does not seem to have been empirically tested yet.
Such a rigorous study is also needed to clarify the parameters of the apparent trade-off between security and usability for a range of fingerprint lengths and provide crucial empirical evidence for designers when deciding on the specifications for key fingerprint verification methods. 

In this work, we contribute to the understanding of the effect of key fingerprint length on the usability and security of manual key fingerprint verification. 
We focus on numeric key fingerprints because of their comparative usability, and specifically consider the \sigwhat\ format, as it is widely deployed.
We present the result of a study in which participants were asked to compare \sigwhat-like key fingerprints of three different lengths.
We measured how the key fingerprint length affects comparison time and accuracy. 
Analyses of our results provide evidence in support of a number of points that so far have been poorly understood in the literature.
Namely, the results show that comparison time only changes significantly when fingerprint pairs of high similarity are being compared, but otherwise stays relatively constant.
Furthermore, we present strong evidence showing that the security non-critical error rate remains fairly low even for long fingerprints, whereas the security critical error rate grows significantly at higher lengths. 
One of the main implications of these results is that \sigwhat\ key fingerprints provide considerably lower levels of security than intended.

\paragraph{Paper outline:}
Section~\ref{sect:Related} summarises the related work, Section~\ref{sect:Design} outlines our research questions and study design, Section~\ref{sect:Results} discusses the results, and Section~\ref{sect:Conclusion} draws conclusions from the study. 

\section{Related work}
\label{sect:Related}

There has been no previous study considering length as an independent variable.
Therefore, in this section we provide a brief overview of the main results on manual fingerprint verification to set out the context in which our work is conducted.

Various formats for key fingerprints have been proposed in the literature or deployed in practice for manual comparison. 
Examples include 
\emph{hexadecimal} e.g.\ GnuPG~\cite{Copeland}, 
\emph{numeric} e.g.\ Signal, WhatsApp, and SafeSlinger~\cite{farb2013safeslinger},
\emph{words} (and pseudo-words) e.g.\ Bubble Babble encoding~\cite{huima2000bubble},
\emph{sentences} e.g.\ pseudo-random poems~\cite{akwizgran2014poem}. 
\emph{graphical} e.g.\ abstract art~\cite{perrig1999hash}, ASCII art~\cite{openssh}, snowflakes~\cite{snowflake}, and unicorns~\cite{Tan2017}, and
\emph{auditory} e.g.\ Loud and Clear~\cite{goodrich2006loud}. 

Several teams investigated the comparative usability of various key representations including Kainda et al.~\cite{kainda09usability}, Dechand et al.~\cite{Dechand2016}, and Tan et al.~\cite{Tan2017}.
These studies broadly found that alphanumeric and numeric representations offer better perceived usability, comparison speed, and accuracy.
The considered numeric fingerprint lengths in these studies were 6, 34, and 48 digits, respectively.

The usability and security of key fingerprint verification for end-to-end encrypted instant messaging apps have been the subject of studies by Herzberg et al.~\cite{Herzberg2016}, Schröder et al.~\cite{Schroder2016}, and Shirvanian et al.~\cite{shirvanian2017pitfalls}.
Evidence presented in these works unanimously points towards high error rates and low perceived usability of manual verification.
More recently, Livsey et al.\ studied word-based manual fingerprint verification when compared visually or verbally and found that visual comparisons are more effective against security non-critical errors~\cite{livsey2021performance}.
Considering the entire authentication ceremony in these apps, Vaziripour et al.\ found low usability, including low completion rates~\cite{vaziripour2017you}. 
Follow-up studies showed rephrasing the task and redesigning the user interface is effective in helping users understand and perform the ceremony correctly~\cite{vaziripour2018action,wu2019something}.

Evidence of low prevalence of manual verification has been reported in the literature.
For instance, in an attempt to study whether users verify SSH key fingerprints, Gutmann approached two large organisations with \say{several thousand computer-literate users}, and found that staff were \say{unable to recall a single case, or locate any records, or any user ever verifying any SSH server key out-of-band}~\cite{Gutmann2011}.

Device pairing methods are related to manual fingerprint verification and have been studied for their comparative usability and security, notably by Kobsa et al.~\cite{kobsa09serial} and Uzun et al.~\cite{uzun2011pairing}.
Comparing numeric fingerprints has been consistently found to be perceived more usable, provide better speed, and lead to less security critical errors compared to other methods in these studies.

A pertinent research question here concerns the most effective adversarial strategy in crafting a similar fingerprint that would pass less attentive human verification.
Cherubini et al.\ provide eye-tracking evidence that attention to compared strings is highest at the beginning of the string and decreases as progress is made towards the end~\cite{CherubiniMCHBH18}.
Furthermore, several works have hypothesised that human attention is heavily biased towards the beginning and end of the compared sequences~\cite{Rieck2002,Gutmann2011,Dechand2016}.

\section{Study Design}
\label{sect:Design}
We consider the \sigwhat\ numeric key fingerprint format because of its comparatively higher usability and its wide deployment. 
As shown in Figure~\ref{img:signal}, these fingerprints are represented in three lines, each containing four 5-digit chunks, in their full format. 
To study the effect of length, we consider three length \emph{conditions}: 
\begin{itemize}
\item 
\textbf{1 Line} (1L): a fingerprint includes \emph{four} 5-digit chunks in 1 line, corresponding to 1 line out of 3 of the full format, 
\item 
\textbf{2 Lines} (2L): a fingerprint includes \emph{eight} 5-digit chunks in 2 lines, corresponding to 2 lines out of 3 of the full format, and
\item 
\textbf{3 Lines} (3L): a fingerprint includes \emph{twelve} 5-digit chunks in 3 lines, corresponding the full \sigwhat\ format. 
\end{itemize}

To minimise the effect of inconsistent formats, we opted for a \emph{between-participants} design with respect to length conditions, i.e.\ each participant will be randomly assigned to one condition and all the fingerprints they compare will be of the same length according to the condition they are assigned. 

Compared key fingerprint pairs can be either matching or non-matching. An adversary may trade off attack success probability with computation and be happy with a nearly matching fingerprint that may fool a proportion of users. To be able to investigate the interplay of the effect of each of these possibilities with that of fingerprint length, we consider three comparison \emph{types}:
\begin{itemize}
    \item \textbf{Safe}: a comparison between a pair of \emph{fully matching} (i.e.\ identical) fingerprints,
    \item \textbf{Adversarial} (Adv.): a comparison between a pair of \emph{nearly matching} fingerprints with only 1 chunk being different, and
    \item \textbf{Random} (Rand.): a comparison between a pair of \emph{randomly selected} (and hence highly dissimilar) fingerprints.
\end{itemize}

The above types represent scenarios where a user encounters an authentic key, an adversarially crafted one in case of an attack, or an erroneous key, respectively. 

To closely follow what would happen in practice where the same user may compare safe, adversarial, or random fingerprints, we opted for a \emph{within-participants} design with respect to comparison types, i.e.\ each participant will carry out comparisons of all types. 

It is expected that in practice users will be comparing safe fingerprints most of the time and the occurrence of attack scenarios will be limited to rare occasions.
Hence, a realistic study should contain as few adversarial pairs as possible. 
At the same time, gathering sufficient data to compute reliable security-critical error rates requires as many adversarial pairs as possible. 
We decided to strike a balance between these two competing goals by designing the study to show \emph{12 safe, 4 adversarial, and 4 random} key fingerprint pairs to each participant. 
Dechand et al.\ follow a similar principle~\cite{Dechand2016}.
The 20 key pairs are shown to the participant in a random order different for each participant to counterbalance the possible effects of habituation and fatigue. 

We emphasise that the scenario we consider is \emph{manual} fingerprint verification carried out \emph{individually}. 
This is also the approach taken by Kainda et al.~\cite{kainda09usability}, Dechand et al.~\cite{Dechand2016}, and Tan et al.~\cite{Tan2017}.
Automated verification, such as scanning the QR code provided by \sigwhat\ using a smartphone camera, and collaborative verification, i.e.\ two users carrying out the comparison together, are both outside the scope of our study. 

\subsection{Adversarial Model}
\label{sect:adv-model}

We consider adversaries that are able to intercept initial key exchange messages between user devices and replace them with adversarially chosen ones. 
However, the adversary does not have the ability to modify messages on the out-of-band channel, i.e.\ the channel through which key fingerprints are compared and verified. 
The goal of the adversary is to impersonate one or both of the entities, corresponding to impersonation or \aitm\ attacks, respectively. 

These capabilities allow the adversary to replace a user's authentic keys with their own which would result in key fingerprints being computed on different keys. 
Specifically for the \sigwhat\ key fingerprint format, we allow adversaries to create key fingerprints that matched all but one of the key fingerprint chunks.
This is to keep the level of similarity high between adversarial pairs.

The \sigwhat\ fingerprint is made of two halves, each a 30-digit fingerprint of the so-called \say{identity key} of one of the two parties~\cite{moxie0-2016-safety-numbers,WhatsAppWhitePaper}.
From each party's viewpoint, the adversary may only compromise one of these two halves since each party \say{knows} the authentic version of their own key.
Hence, we did not allow adversarial digits to cross the midpoint boundary and restricted the adversary to manipulating digits only in the second half of the fingerprint. 
The chunk not targeted for collision by the adversary was designated to be the one just after the key fingerprint midpoint.
This is to maximise the likelihood that it would be overlooked since previous works suggest that users pay less attention to the middle sections of the compared fingerprints~\cite{Rieck2002,Gutmann2011,Dechand2016,CherubiniMCHBH18}.

Requiring all but one of the chunks to be identical in adversarial fingerprints corresponds to \say{adversarial powers} outlined in Table~\ref{table:power} under \say{no iteration} for each condition.
For instance, for our 2~Lines condition, there are eight 5-digit chunks, four of which are computed from the key provided by the adversary. 
The adversary needs three out of these four chunks to be identical to those of the fingerprint half being impersonated, i.e.\ it needs a 3-chunk, i.e.\ 15-digit, collision. 
This is equivalent to finding a second preimage for a hash function with an output length of approximately $49.8$ bits, since $10^{15} \approx 2^{49.8}$. 
Testing every preimage can be seen as a Bernoulli trial and hence the success probability of such an attack with respect to number of computed hashes follows the cumulative distribution function of a \emph{geometric distribution}. 
It follows that the expected number of hashes that need to be computed in the attack is approximately $0.69 \times 2^{49.8}$. 
Despite this, the attack is said to require $2^{49.8}$ adversarial power by convention.

Modern applications use \emph{iterated hashing} for fingerprint calculation to increase the computational cost for adversaries while keeping the cost of hashing for legitimate users within affordable bounds. 
For instance, WhatsApp and Signal iterate the hash 5200 times to compute each fingerprint half.
If such a design is used, the incurred computational cost of attacks will be about $5200 \approx 2^{12.3}$ times higher than the base case where no iteration is used. 
The required adversarial powers, if 5200 iterations are used, are listed in Table~\ref{table:power} under \say{with iteration}. 

\begin{table}
  \centering
  \caption{Adversarial power required to compute attack keys in each condition assuming either no iteration or 5200 iterations}
  \label{table:power}
  \begin{tabular}{l@{\quad}r@{\quad}r@{\quad}r@{\quad}r}
    \toprule
    & \multicolumn{2}{c}{No.\ of Chunks} & \multicolumn{2}{c}{Adversarial Power} \\
    \cmidrule(lr){2-3}\cmidrule(lr){4-5}
    Condition & All & Collision & no iteration & with iteration \\
    \midrule
    1 Line & 4 & 1 & $2^{16.6}$ & $2^{29.0}$ \\
    2 Lines & 8 & 3 & $2^{49.8}$ & $2^{62.2}$ \\
    3 Lines & 12 & 5 & $2^{83.0}$ & $2^{95.4}$ \\
    \bottomrule
  \end{tabular}
\end{table}

We have opted for variable adversarial power to mirror the fact that shorter fingerprints are only appropriate for safer environments, for instance use cases where adversaries are restricted in time or location. An adversary with high power would be able to easily compute keys that lead to full fingerprint collisions for shorter fingerprint lengths which would not allow us to see the effect of similar but not identical fingerprints on user performance.

\subsection{Research Questions}
\label{sect:rq-hyp}

The overall aim of our study is to investigate whether the length and similarity of key fingerprint have significant effects on a person's performance when comparing key fingerprints. 
We focus on user performance in the comparison task, as measured by effectiveness and efficiency.
Perceived usability would be more appropriate for the overall confirmation ceremony and we do not consider it here.
Accordingly, we developed three sets of hypotheses as follows. 

Considering the speed with which participants can compare pairs of key fingerprints as a measure of efficiency, we tested the following high-level hypothesis $H_1^{t \depend \ell}$ on comparison time $t$ with respect to fingerprint length $\ell$, with the alternative hypothesis $H_0^{t \depend \ell}$ defined as the opposite: 

\vspace{1ex}
\fbox{
  \begin{tabular}{@{} l @{\ \ } p{0.78\columnwidth} @{}}
  $H_1^{t \depend \ell}$: & Participants take longer time to compare longer numeric key fingerprints than shorter ones. \\
  \end{tabular}
}
\vspace{1ex}

\noindent
Since we are studying different comparison types, $H_1^{t \depend \ell}$ gives rise to three type-specific hypotheses for safe, adversarial, and random comparisons.

Considering safe, adversarial, and random fingerprint pairs as pairs with maximum, high, and low similarity, we tested the following high-level hypothesis $H_1^{t \depend s}$ on comparison time $t$ with respect to fingerprint similarity $s$, or equivalently comparison type, with the alternative hypothesis $H_0^{t \depend s}$ defined as the opposite: 

\vspace{1ex}
\fbox{
  \begin{tabular}{@{} l @{\ \ } p{0.78\columnwidth} @{}}
  $H_1^{t \depend s}$: & Participants take longer time to compare numeric key fingerprint pairs with higher similarity. \\
  \end{tabular}
}
\vspace{1ex}

\noindent
Similarly, $H_1^{t \depend s}$ is tested at three different fingerprint lengths, giving rise to three length-specific hypotheses.

Considering the accuracy with which participants can compare pairs of key fingerprints as a measure of effectiveness, we tested the following high-level hypothesis $H_1^{e \depend \ell}$ on error rate $e$ with respect to fingerprint length $\ell$, with the alternative hypothesis $H_0^{e \depend \ell}$ defined as the opposite: 

\vspace{1ex}
\fbox{
  \begin{tabular}{@{} l @{\ \ } p{0.78\columnwidth} @{}}
  $H_1^{e \depend \ell}$: & Participants make more mistakes when comparing longer numeric key fingerprints than shorter ones. \\
  \end{tabular}
}
\vspace{1ex}

\noindent
Here, depending on the comparison type we consider, we have two types of errors: 
\begin{itemize}
    \item \textbf{False Acceptance Errors} occur when non-matching fingerprints are incorrectly accepted as matching, and 
    \item \textbf{False Rejection Errors} occur when matching fingerprints are incorrectly rejected as non-matching.
\end{itemize}

\noindent
Consequently, we test two error-type-specific hypotheses, i.e.\ $H_1^{e \depend \ell}$ for false acceptance and false rejection errors.

Note that the security implications of the two types of error can be considerably different. 
False acceptance errors, especially on adversarial fingerprints, would be security-critical as they would allow an \aitm\ attack to go unnoticed. 
However, false rejection errors would only cause inconvenience. 

It is clear that fingerprint length and comparison type are the independent variables, and comparison time, false acceptance and false rejection error rates are the dependent variables in this study. 

\subsection{Ethical Considerations}
\label{sect:Ethics}

The ethical principles of avoidance of harm, informed consent, and data protection were followed throughout the design, data collection, and analysis phases of our study. 
No actual communication channels were attacked. 
Participants were asked for their consent after providing an information sheet at the start of the study.
The participants could withdraw at any time for any reason. 
The information sheet explained the study and that participation was voluntary, and provided the contact details of the investigators. 
No personally identifiable information were collected from participants. 
Only general demographic data were collected to give contextual information. These were age range, gender, highest education level, and presence of a disability. 

A pilot study was used to estimate the time taken to complete the study, based on which we calculated the amount to pay participants in the main study.
We used the living wage for London and New York to ensure that all participants got fair pay for their time.
Participants who withdrew were still paid for their time.
The University of York's Physical Sciences Ethics Committee approved this work before we collected any data.

\subsection{Pilot Study}

First, we ran a pilot study to find any issues in the study design.
We recruited participants locally by offering entry into a raffle for a £25 (GBP) Amazon gift card.
We advertised the pilot study to friends and family on Facebook.

Participants publicly discussed the pilot study on Facebook.
We did not intend this, but it gave us useful insights into how the participants were approaching the pilot study.
Although we did not aim for many piloting participants, we recruited 60 participants, from which we excluded 17 for being inattentive as they indicated that at least one of the random fingerprint pairs matched.
We asked each participant to compare 20 pairs of fingerprints, some identical and some different. 
We made several modifications to our study based on the pilot study feedback as explained below. 

In each individual task, we asked each participant to compare a pair of fingerprints. 
Our question caused confusion for some of our participants, so we reworded the question from \say{Are Alice and Bob's messages safe?} to \say{Do the numbers match?} to make it more clear.
While the original question works well for those familiar with the purpose of key fingerprint verification, it requires a level of knowledge not generally expected of non-experts. 

Some participants were unsure how to proceed, so we added more guidance.
This was especially important as at least one pilot study participant commented \say{it took me waaay [sic] too long to work out it was essentially a \say{compare these numbers} exercise.} 
We showed participants an extra screen before they started which explained the task and showed them where to find the key fingerprint on the screen.
Besides, we added a counter to each page, so that progress through the study was clear to the participants.

\subsection{Main Study}

\begin{figure}
  \centering
  \fcolorbox{gray}{white}{\includegraphics[width=0.97\columnwidth]{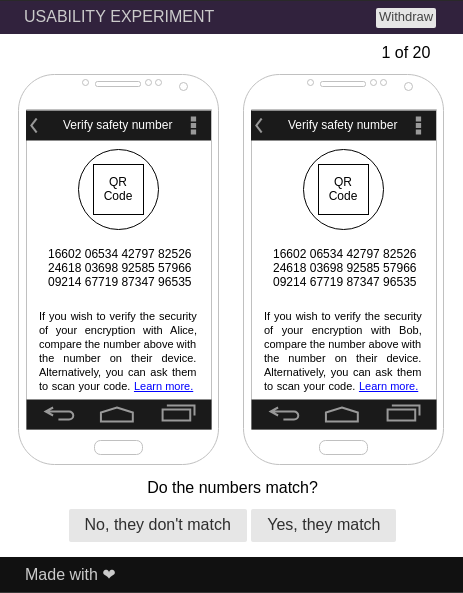}}
  \caption{A screenshot of the study interface for each comparison task as shown to participants.}
  \label{img:experiment}
\end{figure}

In the main study, each participant was randomly assigned one of the three fingerprint lengths, i.e.\ 1, 2, or 3 lines, and asked to compare 20 different key fingerprint pairs of the same length, comprising of 12 safe, 4 adversarial, and 4 random pairs in a randomised order. 
The browser window for each fingerprint pair comparison included two simulated phone screens side-by-side and asked participants \say{Do the numbers match?} with response options \say{Yes, they match} and \say{No, they don't match} as in Figure~\ref{img:experiment}.
Random key fingerprint pairs were used as \emph{attention checkers}.
Participants who got any of the attention checkers wrong were excluded from our analysis, but were still compensated for their time.

Participants were recruited through MTurk.
We did not restrict which MTurk users could accept the task, other than stopping those who had already done the study.
Each participant was paid \$2 (USD) for their time.
All of the guidance was written in English, so all participants needed a sufficient level of English reading comprehension to understand the tasks.
Since the included participants all passed the attention checkers we assume this to be the case.
Before starting the tasks, the participants read the information sheet and consented to take part in the study.

\subsection{Technical Implementation}
We built the experiment on Amazon Web Services (AWS) using Python and TypeScript.
We used AWS Lambda to host the back-end, stored the data encrypted in AWS DynamoDB, and fronted the site with a static site stored in AWS S3 and distributed through AWS CloudFront.
We exposed the Lambda API using AWS API Gateway, which offers TLS by default, so all the participants' data was encrypted in transit. 

\subsection{Study Participants}
A total of 186 participants were recruited. 
2 were excluded from our analysis for failing to complete the study and another 22 for failing the attention checkers. 
In all the following analyses, we report the results for the remaining 162 participants. 
Table~\ref{tab:demographics} shows self-reported participant demographics.
As the table shows, large proportions of our participants declared being male, young, educated, and not disabled.
We had an even split however between conditions: 
53, 55, and 54 participants were assigned to the 1L, 2L, and 3L conditions, respectively. 

\begin{table}
    \centering
    \caption{Participant demographics}
    \begin{tabular}{llrr}
        \toprule
        Demographic & Group & Count & Proportion \\
        \midrule
        Gender 
        & Female & 40 & $\approx$25\% \\
        & Male & 102 & $\approx$63\% \\
        & Other & 1 & $<$1\% \\
        & Preferred not to say & 19 & $\approx$12\% \\
        \midrule
        Age 
        & 18--20 & 3 & $\approx$2\% \\
        & 21--35 & 111 & $\approx$69\% \\ 
        & 36--50 & 29 & $\approx$18\% \\
        & 51--60 & 4 & $\approx$2\% \\
        & 61 and above & 0 & 0\% \\
        & Preferred not to say & 15 & $\approx$9\% \\
        \midrule
        Education 
        & High school diploma & 35 & $\approx$22\% \\
        & Bachelor's degree & 94 & $\approx$58\% \\
        & Master's degree & 17 & $\approx$10\% \\
        & Professional degree & 1 & $<$1\% \\
        & Preferred not to say & 15 & $\approx$9\% \\
        \midrule
        Disability 
        & Declared a disability & 12 & $\approx$7\% \\
        & Declared no disability & 132 & $\approx$81\% \\
        & Preferred not to say & 18 & $\approx$11\% \\
        \bottomrule
    \end{tabular}
    \label{tab:demographics}
\end{table}

\section{Results}
\label{sect:Results}
In this section we give an overview of the collected data and the results of testing the hypotheses stated in Section~\ref{sect:rq-hyp}, using the common $\alpha=0.05$ significance level throughout.

We first tested for any significant demographic difference between groups of users in the three conditions.
Fisher's exact test found no significant difference in the reported gender, educational level, disability, or age between the three groups. The p-values were 0.56, 0.75, 0.91, and 0.28 respectively.

\subsection{Comparison Time}
\begin{figure}
  \centering
  \includegraphics[width=0.9\columnwidth]{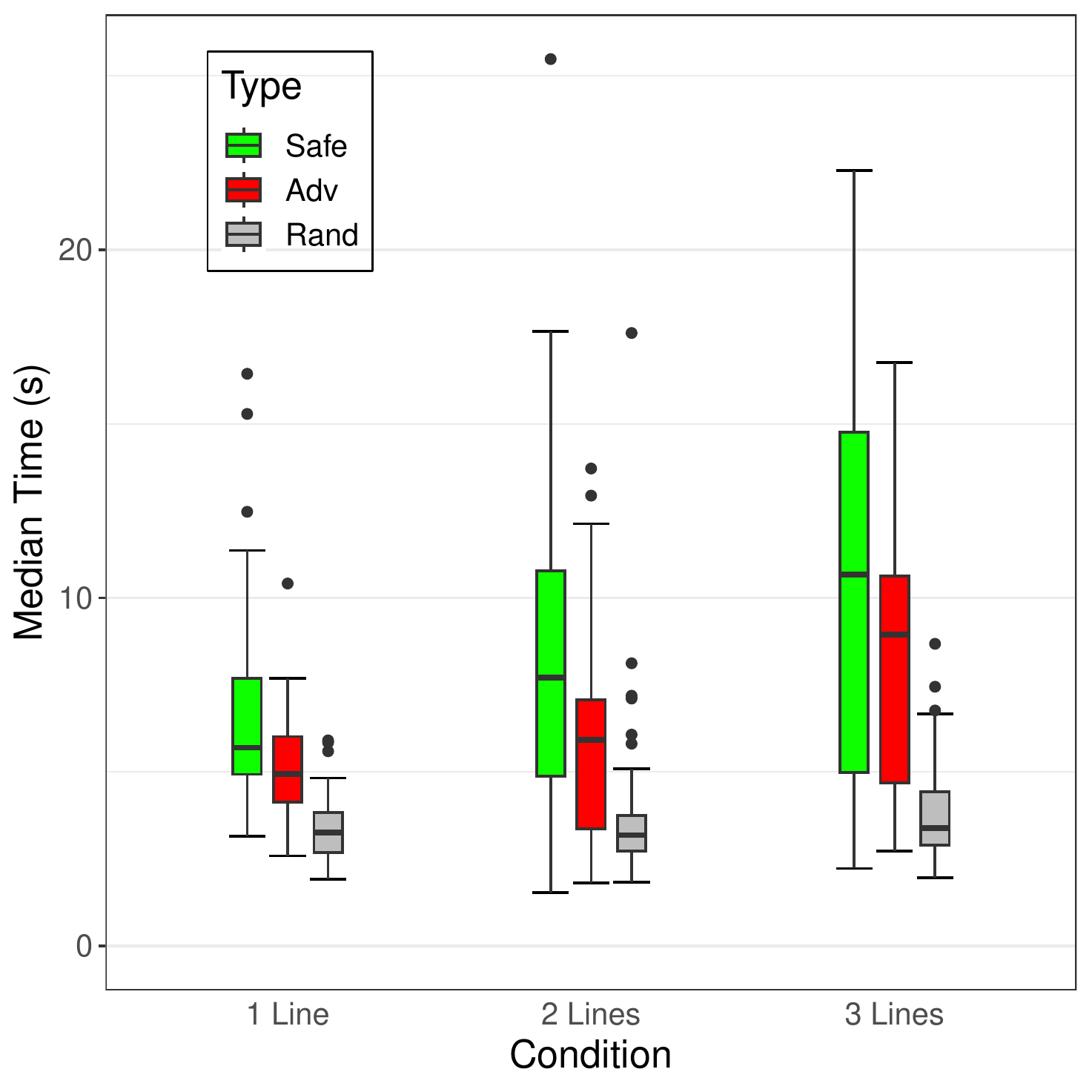}
  \caption{Distributions of participant median times to compare fingerprints by condition (1, 2, 3 Lines) and comparison type (Safe, Adversarial (Adv), Random (Rand))}
  \label{fig:cond-type-time}
\end{figure}

We calculated each participant's median comparison times for each three comparison types: safe, adversarial, and random comparisons. 
The distribution parameters of participant median comparison times by comparison type and condition are detailed in Table~\ref{tbl:time-distros} and depicted in Figure~\ref{fig:cond-type-time}. 
As expected, median comparison times for all nine combinations (3 conditions $\times$ 3 types) have skewed distributions with long tails. 
Shapiro-Wilk tests of normality were significant in all cases except for 1-line adversarial comparisons (1L safe $p<0.001$, adv. $p=0.071$, rand. $p=0.004$, 2L safe $p=0.001$, adv. $p=0.007$, rand. $p<0.001$, 3L safe $p=0.016$, adv. $p=0.028$, rand. $p<0.001$) indicating that 8 out of 9 of the median time distributions are significantly non-normal. 
Hence, non-parametric tests were used for analysis. 
For analysing change with fingerprint length, we have independent samples and hence Kruskal--Wallis test was used, whereas for analysing change with comparison type, we have related measures and hence Friedman test was appropriate.

\begin{table}
  \centering
  \caption{Distribution parameters (lower quartile, \textbf{median}, upper quartile) of median comparison times (in seconds) by comparison type (Safe, Adversarial, and Random) and condition (1, 2, 3 Lines)}
  \label{tbl:time-distros}
  \begin{tabular}{l @{\quad} r@{\ }r@{\ }r @{\quad} r@{\ }r@{\ }r @{\quad} r@{\ }r@{\ }r}
    \toprule 
    Type & \multicolumn{3}{c}{1 Line} & \multicolumn{3}{c}{2 Lines} & \multicolumn{3}{c}{3 Lines} \\
    \midrule 
    Safe & (4.9, & \textbf{5.7}, & 7.7) & (4.9, & \textbf{7.7}, & 10.8) & (5.0, & \textbf{10.7}, & 14.8) \\
    Adversarial & (4.1, & \textbf{4.9}, & 6.0) & (3.4, & \textbf{5.9}, & 7.1) & (4.7, & \textbf{8.9}, & 10.6) \\
    Random & (2.7, & \textbf{3.3}, & 3.8) & (2.7, & \textbf{3.2}, & 3.8) & (2.9, & \textbf{3.4}, & 4.4) \\
    \bottomrule
  \end{tabular}
\end{table}

\subsubsection{Change with Fingerprint Length.}
\label{sect:time-length}
For safe fingerprint comparisons, Kruskal--Wallis test found statistically significant differences between median comparison times for fingerprints of various lengths ($\chi^2(2)=13.3$, $p=0.001$). 
The effect size was moderate ($\eta^2[H]=0.071$).
Pairwise Wilcoxon test between groups with Holm correction found significant differences between all conditions (1L--2L: $W=1112$, $p=0.049$, 2L--3L: $W=1114$, $p=0.049$, 1L--3L: $W=905$, $p=0.003$). 

For adversarial comparisons, Kruskal--Wallis test found statistically significant differences between median comparison times for fingerprints of various lengths ($\chi^2(2)=22.3$, $p < 0.001$).
The effect size was moderate ($\eta^2[H]=0.128$).
Pairwise Wilcoxon test between groups with Holm correction showed that only the differences between 3-line comparisons and the other two groups were significant (1L--2L: $W=1277$, $p=0.269$, 2L--3L: $W=889$, $p<0.001$, 1L--3L: $W=728$, $p<0.001$). 

For random comparisons, Kruskal--Wallis test did not find statistically significant differences between median comparison times for fingerprints of various lengths ($\chi^2(2)=3.68$, $p=0.159$). 

The analysis above shows that although we can reject $H_0^{t \depend \ell}$ for safe comparisons and for adversarial comparisons at higher fingerprint lengths, namely for 2L--3L and 1L--3L comparisons, the same cannot be done for random comparisons. 
This means that our a priori expectation of median comparison time increasing with fingerprint length only holds when similarity between compared fingerprints is high (e.g.\ in the case of safe pairs that are identical), but as the differences between compared fingerprints grow larger (e.g.\ in random pairs) the differences between comparison times for various lengths become insignificant to the point that median comparison times stays approximately constant for 1-line, 2-line, and 3-line random fingerprints.

\subsubsection{Change with Comparison Type.}
\label{sect:time-type}
For 1-line comparisons, Friedman test found statistically significant differences between the distributions of median times for safe, adversarial, and random comparisons ($\chi^2(2)=77.43$, $p < 0.001$). 
The effect size was large (Kendall $W=0.73$). 
Nemenyi post hoc test indicated significant differences between median time distributions for all three pairs of comparison types (safe--adv.: $p=0.001$, adv.--rand.: $p<0.001$, safe--rand.: $p<0.001$).

For 2-line comparisons, Friedman test found statistically significant differences between the distributions of median times for safe, adversarial, and random comparisons ($\chi^2(2)=52.51$, $p < 0.001$). 
The effect size was moderate (Kendall $W=0.48$). 
Nemenyi post hoc test indicated significant differences between median time distributions for all three pairs of comparison types (safe--adv.: $p < 0.001$, adv.--rand.: $p < 0.001$, safe--rand.: $p < 0.001$).

For 3-line comparisons, Friedman test found statistically significant differences between the distributions of median times for safe, adversarial, and random comparisons ($\chi^2(2)=62.11$, $p < 0.001$). 
The effect size was large (Kendall $W=0.58$). 
Nemenyi post hoc test indicated significant differences between median time distributions for all three pairs of comparison types (safe--adv.: $p = 0.011$, adv.--rand.: $p < 0.001$, safe--rand.: $p < 0.001$).

The analysis above shows that for all three different lengths of fingerprints we considered, our participants compare random pairs of fingerprints significantly more quickly than adversarial pairs, and adversarial pairs significantly more quickly than safe pairs. 
Therefore, we emphatically reject $H_0^{t \depend s}$ for all fingerprint lengths.
In other words, the more the differences between the compared fingerprints, the less amount of time it takes on average to compare them and decide whether they are identical or not. 
This observation, coupled with the similar observations in Section~\ref{sect:time-length}, provide considerable evidence supporting the fact that users employ a \say{short-circuit evaluation} like strategy for comparing fingerprints, i.e.\ as soon as a difference is observed a decision is made and the rest of the comparison is abandoned.

\subsection{Error Rates}

In this section we bring the results and analyses of the effect of length on false acceptance and rejection errors. Note that participants who made any errors in comparing random fingerprints were excluded from our study as inattentive participants and hence all attentive participants we consider have correctly identified such fingerprints as non-matching. Consequently, we do not consider random fingerprints in our analysis in this section. We are testing for change with fingerprint length for both error types, hence Kruskal--Wallis was deemed appropriate.

\subsubsection{False Acceptance Errors.}
\label{sect:fae-length}
Each participant in our study carried out 4 adversarial comparisons. 
Table~\ref{table:npart-nfae} lists the number and proportion of participants by number of false acceptance errors they made for different lengths of fingerprints.

\begin{table}
  \centering
  \caption{Number of participants (number/total, top row in each section) and proportion of participants (in bold) including 95\% confidence interval lower and upper limits (bottom row in each section) by number of false acceptance errors out of 4 (denoted by \#) and condition (1, 2, 3 Lines)}
  \label{table:npart-nfae}
  \begin{tabular}{l @{\quad} r@{\ }r@{\ }r @{\quad} r@{\ }r@{\ }r @{\quad} r@{\ }r@{\ }r}
    \toprule 
    \# & \multicolumn{3}{c}{1 Line} & \multicolumn{3}{c}{2 Lines} & \multicolumn{3}{c}{3 Lines} \\
    \midrule 
    0 & \multicolumn{3}{c}{38/53} & \multicolumn{3}{c}{30/55} & \multicolumn{3}{c}{21/54} \\
    & (62\%, & \textbf{72\%}, & 84\%) & (44\%, & \textbf{55\%}, & 69\%) & (26\%, & \textbf{39\%}, & 53\%) \\
    \midrule[0.1\lightrulewidth]
    1 & \multicolumn{3}{c}{8/53} & \multicolumn{3}{c}{7/55} & \multicolumn{3}{c}{8/54} \\
    & (6\%, & \textbf{15\%}, & 28\%) & (2\%, & \textbf{13\%}, & 27\%) & (2\%, & \textbf{15\%}, & 29\%) \\
    \midrule[0.1\lightrulewidth]
    2 & \multicolumn{3}{c}{4/53} & \multicolumn{3}{c}{5/55} & \multicolumn{3}{c}{6/54} \\
    & (0\%, & \textbf{8\%}, & 20\%) & (0\%, & \textbf{9\%}, & 24\%) & (0\%, & \textbf{11\%}, & 25\%) \\
    \midrule[0.1\lightrulewidth]
    3 & \multicolumn{3}{c}{0/53} & \multicolumn{3}{c}{1/55} & \multicolumn{3}{c}{2/54} \\
    & (0\%, & \textbf{0\%}, & 13\%) & (0\%, & \textbf{2\%}, & 16\%) & (0\%, & \textbf{4\%}, & 18\%) \\
    \midrule[0.1\lightrulewidth]
    4 & \multicolumn{3}{c}{3/53} & \multicolumn{3}{c}{12/55} & \multicolumn{3}{c}{17/54} \\
    & (0\%, & \textbf{6\%}, & 18\%) & (11\%, & \textbf{22\%}, & 36\%) & (19\%, & \textbf{31\%}, & 46\%) \\
    \bottomrule
  \end{tabular}
\end{table}

The proportion of participants making no false acceptance error decreases from 72\% for 1-line key fingerprints to 55\% for 2-line fingerprints and eventually to the very low figure of 39\% for 3-line fingerprints which are used by \sigwhat.
On the other hand, while only 6\% of the participants did not manage to spot any of the adversarial comparisons for 1-line fingerprints, this figure rose to 22\% for 2-line fingerprints, and eventually to 31\% for 3-line fingerprints.

Kruskal--Wallis test indicated significant differences between the number of false acceptance errors made by participants for different key fingerprint lengths ($\chi^2(2)=15.03, p<0.001$). 
The effect size was moderate ($\eta^2[H]=0.082$).
Pairwise comparisons using Wilcoxon rank sum test with Holm correction indicated significant differences only between 1-line and 3-line conditions (1L--2L: $p=0.051$, 2L--3L: $p=0.102$, 1L--3L: $p<0.001$).
Therefore we can reject $H_0^{e \depend \ell}$ for false acceptance errors for larger differences between fingerprint lengths.
In other words, we find evidence that indicates false acceptance errors significantly increase when the length of the key fingerprint significantly increases.

To distill these figures, we can compute overall average false acceptance error rates by looking at the number of such errors made over all comparisons across all participants. Since all participants make the same number of adversarial comparisons, this would be equivalent to first computing an average error rate for each participant and then averaging over all participants. Number of false acceptance errors for all participants and their respective rates, including 95\% confidence intervals, are listed in Table~\ref{table:far}. As the figures suggest, an adversary mounting an attack against random users is expected to have an estimated 13.2\% success rate for 1-line, 30.9\% for 2-line, and 43.5\% for 3-line fingerprints.

\begin{table}
  \centering
  \caption{Number of false acceptance errors (error/total, top row) and the \textbf{mean} rate (in bold) including 95\% confidence interval lower and upper limits (bottom row) over all participants by condition (1, 2, 3 Lines)}
  \label{table:far}
  \begin{tabular}{@{} r@{\ }r@{\ }r @{\quad\!} r@{\ }r@{\ }r @{\quad\!} r@{\ }r@{\ }r @{}}
    \toprule 
    \multicolumn{3}{c}{1 Line} & \multicolumn{3}{c}{2 Lines} & \multicolumn{3}{c}{3 Lines} \\
    \midrule 
    \multicolumn{3}{c}{28/212} & \multicolumn{3}{c}{68/220} & \multicolumn{3}{c}{94/216} \\
    (9.0\%, & \textbf{13.2\%}, & 18.5\%) & (24.9\%, & \textbf{30.9\%}, & 37.5\%) & (36.8\%, & \textbf{43.5\%}, & 50.4\%) \\
    \bottomrule
  \end{tabular}
\end{table}

\subsubsection{False Rejection Errors.}
\label{sect:fre-length}
In our study, each participant compared 12 safe (i.e.\ identical) key fingerprints. 
The number and proportion of participants by number of false rejection errors they made for different fingerprint lengths are shown in Table~\ref{table:npart-nfre}. No participant made 7 or above errors and for all categories of 2 to 6 errors, there was at most 1 participant who made that number of errors. We therefore compressed the table for those categories.

\begin{table}
  \centering
  \caption{Number of participants (number/total, top row in each section) and proportion of participants (in bold) including 95\% confidence interval lower and upper limits (bottom row in each section) by number of false rejection errors out of 12 (denoted by \#) and condition (1, 2, 3 Lines)}
  \label{table:npart-nfre}
  \begin{tabular}{@{} l @{\ } r@{\ }r@{\ }r @{\ } r@{\ }r@{\ }r @{\ } r@{\ }r@{\ }r @{}}
    \toprule 
    \# & \multicolumn{3}{c}{1 Line} & \multicolumn{3}{c}{2 Lines} & \multicolumn{3}{c}{3 Lines} \\
    \midrule 
    0 & \multicolumn{3}{c}{49/53} & \multicolumn{3}{c}{47/55} & \multicolumn{3}{c}{43/54} \\
    & (87\%, & \textbf{92\%}, & 98\%) & (78\%, & \textbf{85\%}, & 94\%) & (70\%, & \textbf{80\%}, & 90\%) \\
    \midrule[0.1\lightrulewidth]
    1 & \multicolumn{3}{c}{3/53} & \multicolumn{3}{c}{5/55} & \multicolumn{3}{c}{10/54} \\
    & (0\%, & \textbf{6\%}, & 12\%) & (2\%, & \textbf{9\%}, & 18\%) & (9\%, & \textbf{19\%}, & 29\%) \\
    \midrule[0.1\lightrulewidth]
    2--6 & \multicolumn{3}{c}{0--1/53} & \multicolumn{3}{c}{0--1/55} & \multicolumn{3}{c}{0--1/54} \\
    & (0\%, & \textbf{0--2\%}, & 6--8\%) & (0\%, & \textbf{0--2\%}, & 9--11\%) & (0\%, & \textbf{0--2\%}, & 10--12\%) \\
    \midrule[0.1\lightrulewidth]
    7--12 & \multicolumn{3}{c}{0/53} & \multicolumn{3}{c}{0/55} & \multicolumn{3}{c}{0/54} \\
    & (0\%, & \textbf{0\%}, & 6\%) & (0\%, & \textbf{0\%}, & 9\%) & (0\%, & \textbf{0\%}, & 10\%) \\
    \bottomrule
  \end{tabular}
\end{table}

As the table shows, the proportion of participants making no false rejection errors steadily decreases from 92\% for 1-line fingerprints to 85\% for 2-line fingerprints and eventually to 80\% for 3-line fingerprints. However, the overwhelming majority of participants make no more than 1 error for all fingerprint lengths.

Kruskal--Wallis test did not find significant differences between the number of false rejection errors made by participants for different fingerprint lengths ($\chi^2(2)=3.39, p=0.184$). This shows that although the number of false rejection errors increase with fingerprint length, this increase is not statistically significant for the range of fingerprint lengths we considered and hence we cannot reject $H_0^{e \depend \ell}$ for false rejection rates.

We can again look at the global false rejection error rates over all participants as indicators of the rates with which safe comparisons might be erroneously rejected in general for different fingerprint lengths. These rates are listed in Table~\ref{table:frr} and show that false rejection errors are rare, with upper confidence limits of less than 5\% for all fingerprint lengths. 
Besides, there does not seem to be a considerable change in error rates as fingerprints get longer, especially at higher lengths. 

\begin{table}
  \centering
  \caption{Number of false rejection errors (error/total, top row) and the \textbf{mean} rate (in bold) including 95\% confidence interval lower and upper limits (bottom row) over all participants by condition (1, 2, 3 Lines)}
  \label{table:frr}
  \begin{tabular}{r@{\ }r@{\ }r @{\quad} r@{\ }r@{\ }r @{\quad} r@{\ }r@{\ }r}
    \toprule 
    \multicolumn{3}{c}{1 Line} & \multicolumn{3}{c}{2 Lines} & \multicolumn{3}{c}{3 Lines} \\
    \midrule 
    \multicolumn{3}{c}{6/636} & \multicolumn{3}{c}{18/660} & \multicolumn{3}{c}{13/648} \\
    (0.3\%, & \textbf{0.9\%}, & 2.0\%) & (1.6\%, & \textbf{2.7\%}, & 4.3\%) & (1.1\%, & \textbf{2.0\%}, & 3.4\%) \\
    \bottomrule
  \end{tabular}
\end{table}

\subsection{Comparison with Previous Work}

To put our results in context, in this section we list the comparison times and error rates reported in previous studies on numeric fingerprint verification alongside our results. 
These measurements are not directly comparable per se, since they are collected under different conditions. 
Nevertheless, we believe this comparison helps situate our results in the wider context. 

\begin{figure}
    \centering
    \includegraphics[width=0.9\columnwidth]{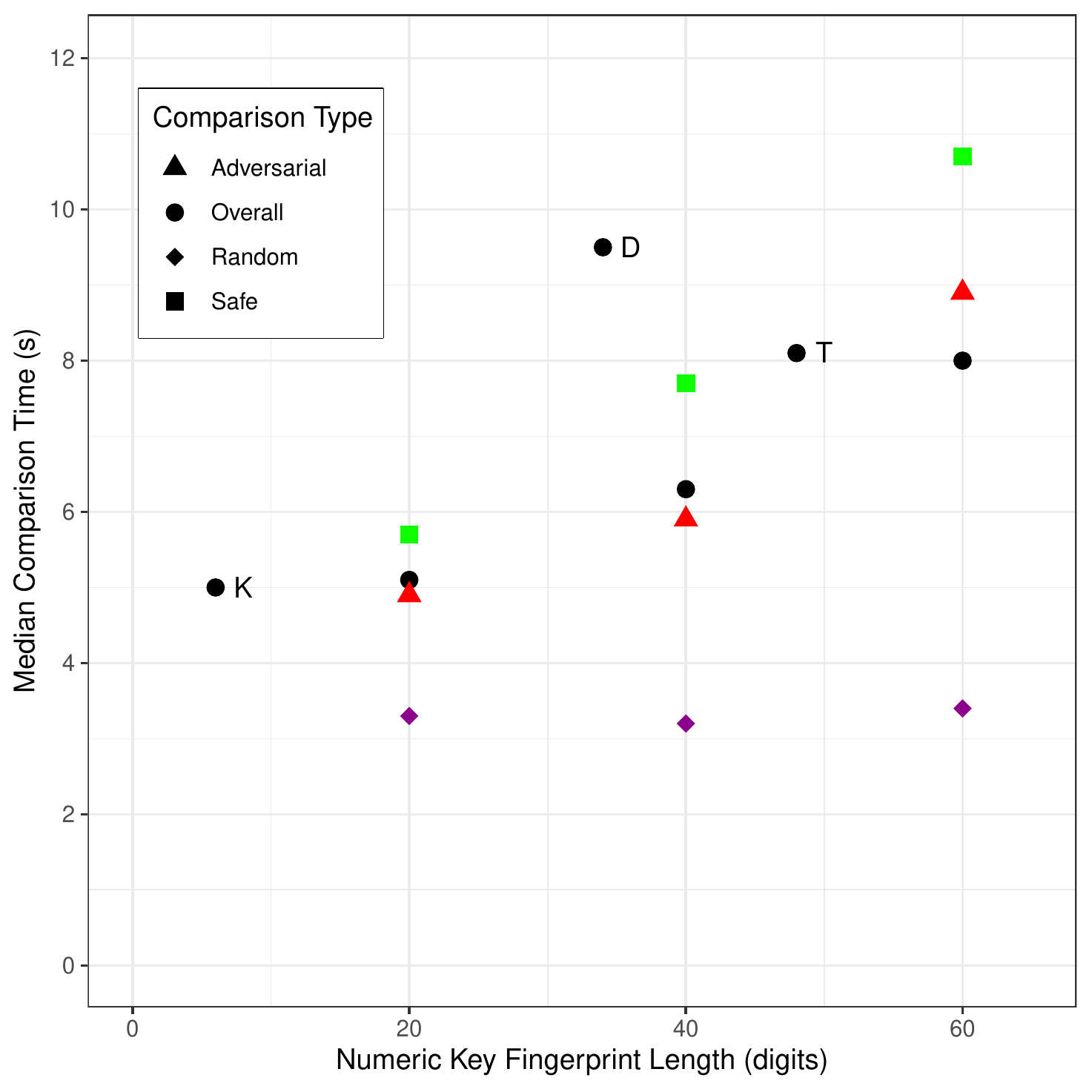}
    \caption{Median comparison times measured in our study (at lengths 20, 40, 60 digits) compared to those reported in the literature (annotated K: Kainda et al.~\cite{kainda09usability}, D: Dechand et al.~\cite{Dechand2016}, T: Tan et al.~\cite{Tan2017})}
    \label{fig:time-comp}
\end{figure}

Results show a gradual increase of comparison time with fingerprint length as expected. 
Kainda et al.\ reported a median of 5 and a mean of 6 seconds, respectively, for comparing 6-digit numeric fingerprints~\cite{kainda09usability}. 
Other notable results are a median of 9.5 seconds for 34-digit fingerprints reported by Dechand et al.~\cite{Dechand2016} and a median of 8.1 seconds for 48-digit fingerprints by Tan et al.~\cite{Tan2017}. 
These works all only report overall results and do not give a breakdown of the results by type, i.e.\ safe, random, and adversarial comparisons.
The overall medians in our study can be computed as 5.1, 6.3, and 8.0 seconds for 20, 40, and 60-digit fingerprints (i.e.\ for 1, 2, and 3-line fingerprints), and the respective means as 6.1, 7.4, and 10.0 seconds.
Overall median comparison times for our study and the previous studies are all shown in Figure~\ref{fig:time-comp}. 
We have also included median comparison times for the three comparison types in our study, but excluded Uzun et al.'s reported mean of 12.5 seconds as it was for a pair of users carrying out the comparison collaboratively~\cite{uzun2011pairing}. 

As the figure shows, our overall results and those of Kainda et al. and Tan et al. are more or less in line with each other, with Dechand et al.'s result seemingly being an outlier to some extent.
Another important point depicted by our results is that overall medians only give reliable estimates in environments where occasional attacks and random comparisons are expected. In safer environments, where the overwhelming majority of the comparisons are expected to be safe ones, timing estimates should be considered to be considerably higher, e.g., by about a third for 60-digit fingerprints. 

\begin{figure}
    \centering
    \includegraphics[width=0.9\columnwidth]{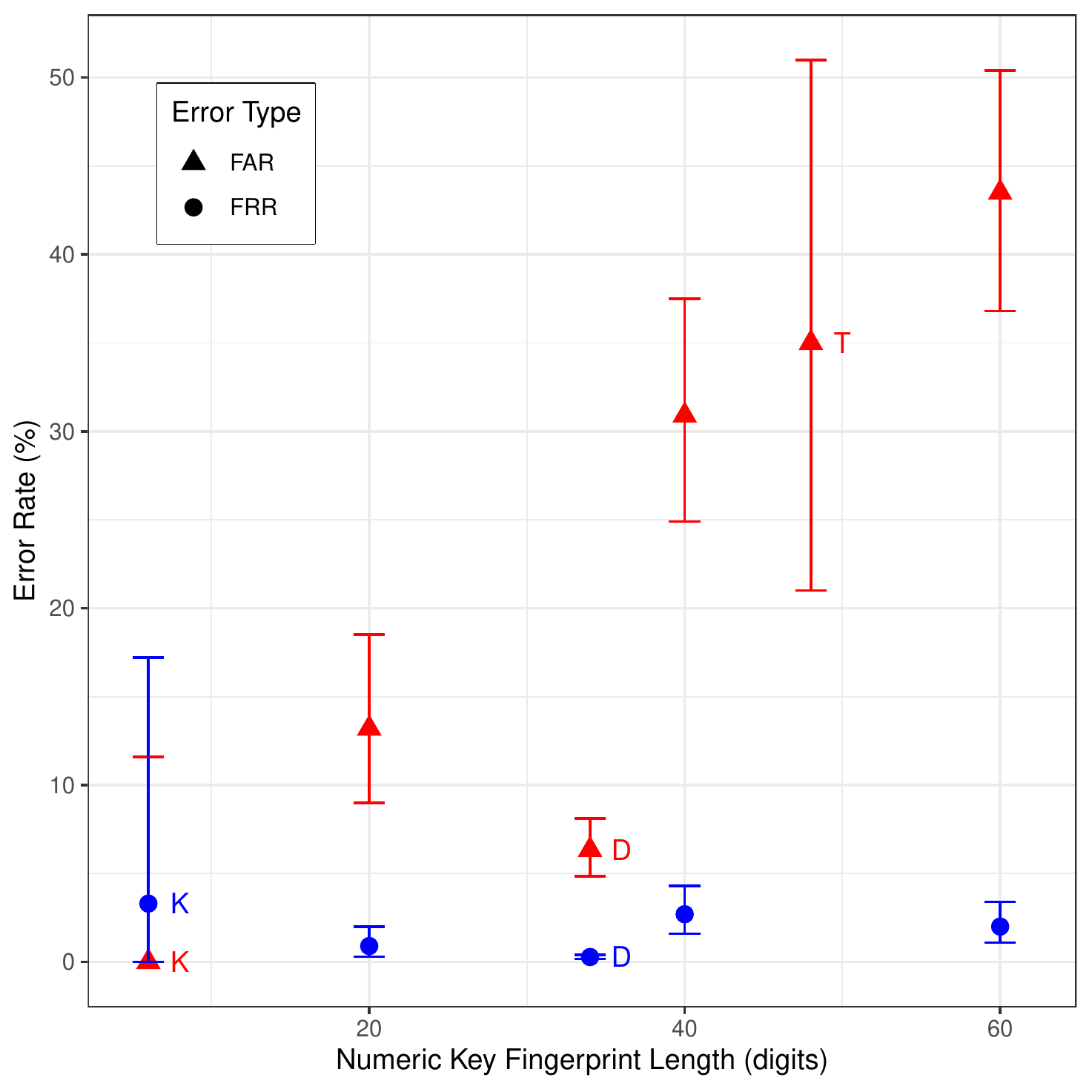}
    \caption{Average false acceptance and rejection rates (FAR, FRR) and 95\% confidence intervals measured in our study (at lengths 20, 40, 60 digits) compared to those reported in the literature (annotated K: Kainda et al.~\cite{kainda09usability}, D: Dechand et al.~\cite{Dechand2016}, T: Tan et al.~\cite{Tan2017})}
    \label{fig:far-frr-comp}
\end{figure}

Kainda et al.\ did not observe any false acceptance errors (called \say{security failure} there) in their 30-participant study for 6-digit fingerprints~\cite{kainda09usability}. 
Dechand et al.\ reported a 6.3\% rate (called \say{fail rate}) for 34-digit fingerprints~\cite{Dechand2016} and Tan et al.\ a 35\% rate (called \say{fraction [of attacks] missed}) for 48-digit fingerprints~\cite{Tan2017}. 
Our results of 13.2\%, 30.9\%, and 43.5\% false acceptance error rates for 20, 40, and 60-digit fingerprints are broadly in line with the results above, except for that of Dechand et al.'s, as shown in Figure~\ref{fig:far-frr-comp}. 
A possible explanation for the discrepancy between Dechand et al.'s result and the rest, both in terms of comparison time and error rates, is that Dechand et al.'s participants were particularly attentive and hence took longer time for carrying out the comparisons, ending up with much lower error rates.

As for false rejection rates, Kainda et al.\ report a rate of 3.3\% (called \say{non-security failure}) for 6-digit fingerprints~\cite{kainda09usability} and Dechand et al.\ 0.28\% (called \say{false positive}) for 34-digit fingerprints~\cite{Dechand2016}. 
Tan et al.\ do not report the rate.
Our rates of 0.9\%, 2.7\%, and 2.0\% for 20, 40, and 60-digit fingerprints are largely consistent with the results above.
As Figure~\ref{fig:far-frr-comp} shows, mean false rejection rate remains below 5\% irrespective of the length of compared fingerprints.


\subsection{Limitations}

It is not immediately clear what the best method is to control the similarity between pairs of fingerprints, ensuring adversarial pairs of different lengths have comparable similarity. 
For numeric fingerprints represented without chunking and in one line, one may keep the proportion of different digits constant for various fingerprint lengths.
However when chunking and multiple lines come into play, factors such as where in each line and between chunks the differences appear and how many chunks are affected need to be taken into account. 
We aimed for a simple method of allowing one chunk of difference for all lengths, but this would mean that the proportion of different digits will not stay the same.

In our adversarial comparisons, we considered near-collision fingerprints differing only in one chunk immediately after the midpoint. 
This means that the non-identical chunk appeared in different positions in different conditions: in the middle of the line for the 1 Line condition, in the beginning of the second line for the 2 Line condition, and in the middle of the middle line for the 3 Line condition.
This may have introduced a confound in comparing the 2 Line condition results with the other two conditions, but the comparisons between 1 Line and 3 Line conditions are not expected to have been affected.

We have simulated smartphone user interfaces within browsers. 
In practice, comparisons are made on two real smartphones that are likely to be different makes or models.
However, we don't expect this issue to have had a considerable effect on our results in general.

Our participants were largely young (around 69\% 21--35), male (around 63\% male), and educated (around 69\% with tertiary education). 
This needs to be kept in mind when considering the results.

\section{Discussions and Conclusions}
\label{sect:Conclusion}
We discuss the implications of our results and some possible directions for future work in this section.

\subsection{Implications of the Results}
\begin{figure}
    \centering
    \includegraphics[width=0.9\columnwidth]{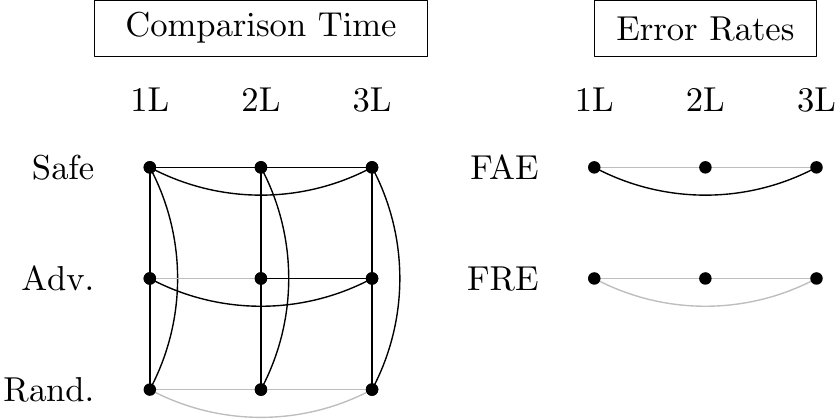}
    \caption{Summary of statistical significance results for comparison time by condition (1L, 2L, 3L) and type (Safe, Adv., Rand.) on the left, and for error rates (FAE, FRE) by condition (1L, 2L, 3L) on the right. Black lines indicate statistically significant differences and grey lines indicate non-significance.}
    \label{fig:significance}
\end{figure}

Figure~\ref{fig:significance} shows the summary of our results in terms of statistical significance for comparison time on the left and error rates on the right.
We list the main takeaways from our study based on the results in the following.
Although these are not mutually exclusive, it is instructive to look at the results from various perspectives.

\paragraph{Fingerprint length is a major determinant of efficiency.}
As the analysis in Section~\ref{sect:time-length} shows, for safe comparisons, changes in comparison time are significant with respect to fingerprint length for all length differences.
In the most common use cases of numeric key fingerprint verification, the overwhelming majority of comparisons are expected to be safe comparisons.
Hence, our results provide strong evidence for the intuition that fingerprint length should be considered as a significant determinant of efficiency when designing numeric key fingerprint verification systems.

\paragraph{Overall time estimates can be misleading.}
Analysis in Section~\ref{sect:time-type} demonstrates that time differences between comparison types are significant at all lengths, with timing estimates for safe comparisons being significantly higher than other types.
Given that in most common use cases we expect safe comparisons to dominate, median comparisons times in practice are going to be closer to median safe comparison times.
However, overall comparison times usually reported in the literature assume arbitrary and unrealistic proportions of safe, adversarial, and random comparisons.
Hence, when considering efficiency, design decisions for common use cases should be made based on safe comparison times, when available, rather than overall comparison times usually reported in the literature.
If safe comparison times are not available, our results show they can be estimated to be between a tenth to a third above overall times depending on fingerprint length.

\paragraph{Users are neither efficient nor effective in comparing highly similar long fingerprints.}
Focusing on adversarial fingerprints with high similarity, the results in Sections~\ref{sect:time-length} and \ref{sect:fae-length} show that although users take significantly longer time to perform the comparison, they make significantly higher false acceptance errors which can be security critical. 
This underlines the crucial role of providing alternative or complimentary means of key fingerprint verification for contexts where higher levels of security is required, as manual verification of long fingerprints suffers from low usability.

\paragraph{Manual key fingerprint verification provides a lower security level than usually assumed.}
Fingerprint lengths are usually chosen to provide desired levels of security.
This level of security indicates the adversarial power required to achieve a (full) fingerprint \emph{collision} (i.e.\ an adversarial fingerprint identical to an authentic one) and hence fool the user with a success probability of 1. 
For instance, the \sigwhat\ fingerprint is designed to provide 112-bit security since the adversarial power required for finding a second preimage for 30-digit key fingerprints computed with 5200 hash iterations is $10^{30} \times 5200 \approx 2^{99.7} \times 2^{12.3} \approx 2^{112}$. 
This means that with approximately $0.69 \times 2^{112}$ hash computations, an adversary is expected to achieve a 50\% success rate. 
Looking at another point of interest on the attack success probability curve (specified in Section~\ref{sect:adv-model}), to achieve a 40\% attack success rate, the adversary would be expected to perform approximately $0.51 \times 2^{112} \approx 2^{111}$ computations.
However, as our results in Section~\ref{sect:fae-length} show, a \emph{near collision} (i.e.\ an adversarial fingerprint sufficiently similar to an authentic one) is enough to achieve a considerable false acceptance error rate as high as 40\%. 
As Table~\ref{table:power} shows, such a near collision would only require $2^{95.4}$ adversarial power, i.e.\ approximately $0.69 \times 2^{95.4} \approx 2^{94.9}$ hash computations.
False acceptance error rate is strongly indicative of the success rate for attack campaigns targeting multiple victims repeatedly, which can be possible in many use cases of such fingerprints.
Hence, it is more realistic to think of the \sigwhat\ fingerprint length providing approximately 96-bit security rather than 112-bit security, 
and in general, longer fingerprint lengths for which high false acceptance rates are possible should be considered to provide considerably less security than usually assumed. 

\paragraph{Users are quite efficient and effective in recognising dissimilar fingerprints.}
Our results for random fingerprint comparisons in Sections~\ref{sect:time-type} and \ref{sect:fre-length} clearly show that not only users are pretty quick and accurate in recognising highly dissimilar fingerprint pairs, but also both comparison time and false rejection error rate stay low and roughly constant even with considerable changes in fingerprint length.
As discussed before, this points toward a \say{short-circuit evaluation} like behaviour exhibited by users in performing fingerprint comparison.
Consequently, in an environment where users may be expected to perform higher proportion of such comparisons, designers can be confident that users can handle a wide range of fingerprint lengths with similar effectiveness and efficiency.

\paragraph{False rejection errors are rare.}
False rejection errors and rates stay quite low across a relatively wide range of fingerprint lengths as the results in Section~\ref{sect:fre-length} show. Indeed, even the 95\% confidence interval upper limits stay below 5\% in all the measurements we carried out.
Therefore, when designing such mechanisms, decisions for fingerprint length can be made mainly based on efficiency and security (including false acceptance errors).

\paragraph{Similarity is a significant determinant of efficiency.}
The most emphatic results were given by the analysis of comparison time with respect to comparison type in Section~\ref{sect:time-type}: the differences between comparison time for safe and adversarial, as well as between adversarial and random, and hence between safe and random fingerprint pairs are found to be significant at all lengths.
This shows that the effect of similarity between fingerprints is markedly significant on the efficiency of manual key fingerprint comparison.

\subsection{Future Work}
As with any other study, the scope of the parameters had to be limited in our investigation and further work is required to explore the parameter space more broadly.
Of particular interest would be investigating a higher granularity of lengths and a wider range of similarity between fingerprints.

To test whether our results can be generalised to wider contexts, it would be crucial to replicate the investigation for other verification modes, including verbal and collaborative comparisons, and other fingerprint representations, including word-based ones.

Our results can be seen as part of a series of related works collectively demonstrating the poor usability of currently recommended methods for manual verification of long key fingerprints, e.g.\ those used by \sigwhat, and underlining the importance of developing better manual and automated verification methods. 

\subsection*{Acknowledgement}
We sincerely thank the reviewers of ARES'23 for their valuable and constructive comments.

\newpage
\bibliographystyle{ACM-Reference-Format}
\bibliography{citations}

\end{document}